\documentclass[12pt,preprint]{aastex}

\begin{document}

\title{3D Relativistic Magnetohydrodynamic Simulations of Magnetized
Spine-Sheath Relativistic Jets}

\author{
Yosuke Mizuno\altaffilmark{1,4}, Philip Hardee\altaffilmark{2} and Ken-Ichi
Nishikawa\altaffilmark{1,3}}

\altaffiltext{1}{National Space Science and Technology Center, 320
Sparkman Drive, VP 62, Huntsville, AL 35805, USA;
Yosuke.Mizuno@msfc.nasa.gov}
\altaffiltext{2}{Department of
Physics and Astronomy, The University of Alabama, Tuscaloosa, AL
35487, USA}
\altaffiltext{3}{Center for Space
Plasma and Aeronomic Research, University of Alabama in Huntsville
, Huntsville, AL 35899, USA}
\altaffiltext{4}{NASA Postdoctoral
Program Fellow/ NASA Marshall Space Flight Center}

\shorttitle{3D RMHD simulations of spine-sheath jets}
\shortauthors{Mizuno et al.}

\begin{abstract}

Numerical simulations of weakly magnetized and strongly magnetized
relativistic jets embedded in a weakly magnetized and strongly
magnetized stationary or weakly relativistic ($v = c/2$) sheath have
been performed. A magnetic field parallel to the flow is used in these
simulations performed by the new GRMHD numerical code RAISHIN used in
its RMHD configuration.  In the numerical simulations the Lorentz
factor $\gamma = 2.5$ jet is precessed to break the initial
equilibrium configuration.  In the simulations sound speeds are
$\lesssim c/\sqrt 3$ in the weakly magnetized simulations and
$\lesssim 0.3~c$ in the strongly magnetized simulations. The
Alfv\'en wave speed is $\lesssim 0.07~c$ in the weakly magnetized
simulations and $\lesssim 0.56~c$ in the strongly magnetized
simulations. The results of the numerical simulations are compared to
theoretical predictions from a normal mode analysis of the linearized
relativistic magnetohydrodynamic (RMHD) equations capable of
describing a uniform axially magnetized cylindrical relativistic jet
embedded in a uniform axially magnetized relativistically moving
sheath. The theoretical dispersion relation allows investigation of
effects associated with maximum possible sound speeds, Alfv\'{e}n wave
speeds near light speed and relativistic sheath speeds.  The
prediction of increased stability of the weakly magnetized
system resulting from $c/2$ sheath speeds and the stabilization of the
strongly magnetized system resulting from $c/2$ sheath speeds is
verified by the numerical simulation results.

\end{abstract}


\section{Introduction}

Relativistic jets are associated with galaxies and quasars ({\bf
AGN}), with black hole binary star systems, and are thought
responsible for the gamma-ray bursts ({\bf GRBs}).  In AGN and
microquasar jets proper motions of intensity enhancements indicate
motions that are mildly superluminal for the microquasar jets $\sim
1.2~c$ (Mirabel \& Rodriquez 1999), and range from subluminal ($\ll
c$) to superluminal ($ \lesssim 6~c$) along the M\,87 jet (Biretta et al.
 1995, 1999), up to $\sim 25~c$
along the 3C\,345 jet (Zensus et al. 1995; Steffen et al.\
1995), and have inferred Lorentz factors $\gamma > 100$ in the GRBs
(e.g., Piran 2005).  The various observed proper motions along AGN and
microquasar jets imply speeds from $\sim 0.9~c$ up to $\sim 0.999~c$,
and the inferred speeds for the GRBs are $\sim 0.99999~c$.

Jets at the larger scales may be kinetically dominated and contain
relatively weak magnetic fields, e.g., equipartition between magnetic
and gas pressure or less, but the possibility of much stronger
magnetic fields certainly exists closer to the acceleration and
collimation region. Here general relativistic magnetohydrodynamic
(GRMHD) simulations of jet formation (e.g., Koide et al.\ 1998, 1999,
2000; Nishikawa et al.\ 2005; De Villiers et al.\ 2003, 2005; Hawley
\& Krolik 2006; McKinney \& Gammie 2004; McKinney 2006; Mizuno et al.\
2006b) and earlier theoretical work (e.g., Lovelace 1976; Blandford
1976; Blandford \& Znajek 1977; Blandford \& Payne 1982) invoke strong
magnetic fields.  Additionally, Vlahakis and Konigl have argued that
magnetically dominated ``Poynting flux'' jets could produce the
accelerations observed in AGN jets such as that in NGC\,6251 and
3C\,345 (Vlahakis \& Konigl 2004) or provide the impetus for high
Lorentz factor gamma-ray bursts outflows (Vlahakis \& Konigl 2003).
In these cases acceleration occurs up to the point at which Poynting
fluxes and kinetic energy fluxes become comparable.  In addition to
strong magnetic fields, a number of GRMHD simulation studies of jet
formation indicate that highly collimated high speed jets driven by
the magnetic fields threading the ergosphere may themselves reside
within a broader wind or sheath outflow driven by the magnetic fields
anchored in the accretion disk (e.g., McKinney 2006; Hawley \& Krolik
2006; Mizuno et al.\ 2006b). This configuration might additionally be
surrounded by a less collimated accretion disk wind from the hot
corona (e.g., Nishikawa et al.\ 2005).

Recent observations of high speed winds in several QSO's with speeds,
$\sim 0.1 - 0.4c$, also indicate that a highly relativistic jet could
reside in a high speed wind or sheath, at least close to the central
engine (Chartas et al.\ 2002, 2003; Pounds et al.\ 2003a, 2003b;
Reeves et al.\ 2003).  For some time other observational evidence such
as {\it limb brightening} has been interpreted as evidence for a
slower external flow surrounding a faster jet spine, e.g., Mkn\,501
(Giroletti et al.\ 2004), M\,87 (Perlman et al. 2001), and a few other
radio galaxy jets (e.g., Swain et al.\ 1998; Giovannini et al.\ 2001).
Additional circumstantial evidence such as the requirement for large
Lorentz factors suggested by the TeV BL Lacs when contrasted with much
slower observed motions has been used to suggest the presence of a
spine-sheath morphology (Ghisellini et al.\ 2005).  Siemignowska et
al.\ (2006) have proposed a two component (spine-sheath) model to
explain the broad-band emission from the PKS 1127-145 jet.  Additionally, a
spine-sheath jet structure has been proposed based on theoretical
arguments (e.g., Sol et al.\ 1989; Henri \& Pelletier 1991; Laing 1996;
Meier 2003) and similar type structure has been investigated in the
context of GRB jets (e.g., Rossi et al.\ 2002; Lazzatti \& Begelman
2005; Zhang et al.\ 2003, 2004; Morsony et al.\ 2006).

In most jet generation numerical and theoretical work it has been
necessary to assume an axisymmetric configuration.  Given the
helicity in the real system, e.g., helical magnetic field or
outwards flow combined with rotation, there exists a potential
problem with the stability of the system.  Obviously, at least for
most AGN jets, stability problems are surmounted and a highly
collimated relatively stable flow is produced.  In this paper we
begin a 3D numerical study of the stability properties of highly
relativistic jet flows allowing for the effects of strong magnetic
fields and relativistic flow in a sheath around the highly
relativistic jet. We note that observed relatively stable jet flow
along with observed jet structures might then be used to constrain
the configuration in the acceleration and collimation region where
magnetic field strengths are high.

In the past, 3D numerical simulations of relativistic unmagnetized jets
along with accompanying theoretical work (e.g., Hardee et al.\ 2001;
Agudo et al.\ 2001; Hardee \& Hughes 2003) has provided relatively
unambiguous interpretation and understanding of structures observed in
the numerical simulations.  Thus, we begin our study by adopting a
simple system, no radial dependence of quantities inside the jet and
also no radial dependence of quantities outside the jet.  This ``top
hat'' configuration can be described exactly by the linearized RMHD
equations.  In general, the system consisting of a jet with ``top
hat'' profile and magnetic field parallel to the flow along with a
uniform external medium also with magnetic field parallel to the flow
is more stable than a system with magnetic and flow helicity.  Such a
system is stable to current driven ({\bf CD}) modes of instability
(Istomin \& Pariev 1994, 1996; Lyubarskii 1999) but can be unstable to
Kelvin-Helmholtz ({\bf KH}) modes of instability (Hardee 2004). This
approach allows us to look at the potential KH modes without
complications arising from coexisting CD modes (see Baty et al.\ 2004)

This paper is organized as follows. In section 2, we describe the
numerical simulation setup and present the results of the
three-dimensional RMHD simulations of spine-sheath relativistic
jets. In section 3, we present the theoretical dispersion relation
that arises from a normal mode analysis of the linearized RMHD
equations, present relevant analytical approximate solutions,
numerically solve the dispersion relation for the simulation
parameters and compare to the simulation results. In section 4 we
conclude.

\section{Numerical Simulations of Spine-Sheath Jets}

\subsection{Numerical Method}

In order to study the long-term stability of magnetized sheath-spine
relativistic jets, we use the 3-dimensional GRMHD code ``RAISHIN''
with Cartesian coordinates in special relativity. The method is based
on a 3+1 formalism of the general relativistic conservation laws of
particle number and energy momentum, Maxwell equations, and Ohm's law
with no electrical resistance (ideal MHD condition) in a curved
spacetime (Mizuno et al.\ 2006a). The RAISHIN code performs special
relativistic calculations in Minkowski spacetime by changing the
metric.

In the RAISHIN code, a conservative, high-resolution shock-capturing
scheme is employed. The numerical fluxes are calculated using the
Harten, Lax, \& van Leer (HLL) approximate Riemann solver scheme. The
flux-interpolated, constrained transport scheme is used to maintain a
divergence-free magnetic field. The RAISHIN code has proven to be accurate
to second order and has passed a number of numerical tests including highly
relativistic cases, and highly magnetized cases in both special and
general relativity (Mizuno et al.\ 2006a).

In the simulations a ``preexisting'' jet flow is established across
the computational domain. This setup represents the case in which the
jet is in equilibrium with an external medium far behind the leading
edge Mach disk and bow shock. We allow the jet flow to be surrounded
by a lower-density external magnetized wind medium. For all
simulations, the ratio of densities is $\rho_{j}/\rho_{e}=2.0$, where
$\rho$ is the mass density in the proper frame. The jet flow has
$u_{j}=0.9165c$ and $\gamma \equiv (1 - u^{2})^{-1/2}=2.5$. The
initial magnetic field is assumed to be uniform and parallel to the
jet flow. The jet is established in static total pressure balance with
the external magnetized wind medium.
Our choice of colder jet in a hotter wind is representative of a jet
spine in a hotter sheath or cocoon as might occur as a result of
astrophysical jet interaction with the surrounding medium. However, the
specific parameters have been chosen for numerical and theoretical
comparison convenience.

The computational domain is $6 R_{j} \times 6 R_{j} \times 60 R_{j}$
with $60 \times 60 \times 600$ computational zones (10 computational
zones span $R_{j}$).  We impose outflow boundary conditions on all
surfaces except the inflow plane at $z = 0$.

A precessional perturbation is applied at the inflow by imposing a
transverse component of velocity with
$u_{\bot}=0.01u_{j}$. Simulations have been performed with
precessional perturbations of angular frequency $\omega
R_{j}/u_{j}=0.40$ (simulation A, $\omega 1$), $0.93$ (simulation B, $\omega 2$)
and $2.69$ (simulation C, $\omega 3$). The simulations are halted after
$\sim 60$ light
crossing times of the jet radius, before the perturbation has crossed
the entire computational domain.

We have performed two sets of simulations. RHD cases are weakly
magnetized (see Table 1). The relevant sound speeds are $a_{e}=0.574c$
and $a_{j}=0.511c$, where the sound speed $a$ is given by
\begin{equation}
a \equiv \left[ {\Gamma p \over \rho + (\Gamma / \Gamma-1) (p/c^{2})}
\right]^{1/2}~,
\end{equation}
with $\Gamma = 13/6$ as the adiabatic index appropriate to a mixture
of relativistically hot electrons and cold baryons (Synge 1957).
The relevant Alfv\'{e}n speeds are $v_{Ae}=0.0682c$ and
$v_{Aj}=0.064c$ , where the Alfv\'en speed $v_{A}$ is given by
\begin{equation}
v_{A} \equiv \left[ {B^{2} \over \rho + (\Gamma / \Gamma-1) (p/c^{2})
+ B^{2}} \right]^{1/2}~.
\end{equation}
Therefore the Alfv\'{e}n speed is much smaller than sound speed. RMHD
cases are strongly magnetized (see Table 1). The relevant
sound speeds are $a_{e}=0.30c$ and $a_{j}=0.226c$. The relevant
Alfv\'{e}n speeds are $v_{Ae}=0.56c$ and $v_{Aj}=0.45c$. In this case the
Alfv\'{e}n speeds are approximately twice the sound speeds. In order to
investigate the effect of an external wind, we have performed a no
wind case ($u_{e}=0$, simulation `n') and a mildly
relativistic wind case ($u_{e}=0.5~c$, simulation `w').

\placetable{table 1}

\subsection{Numerical Results}

\begin{figure}[hp!]
\epsscale{0.8}
\plotone{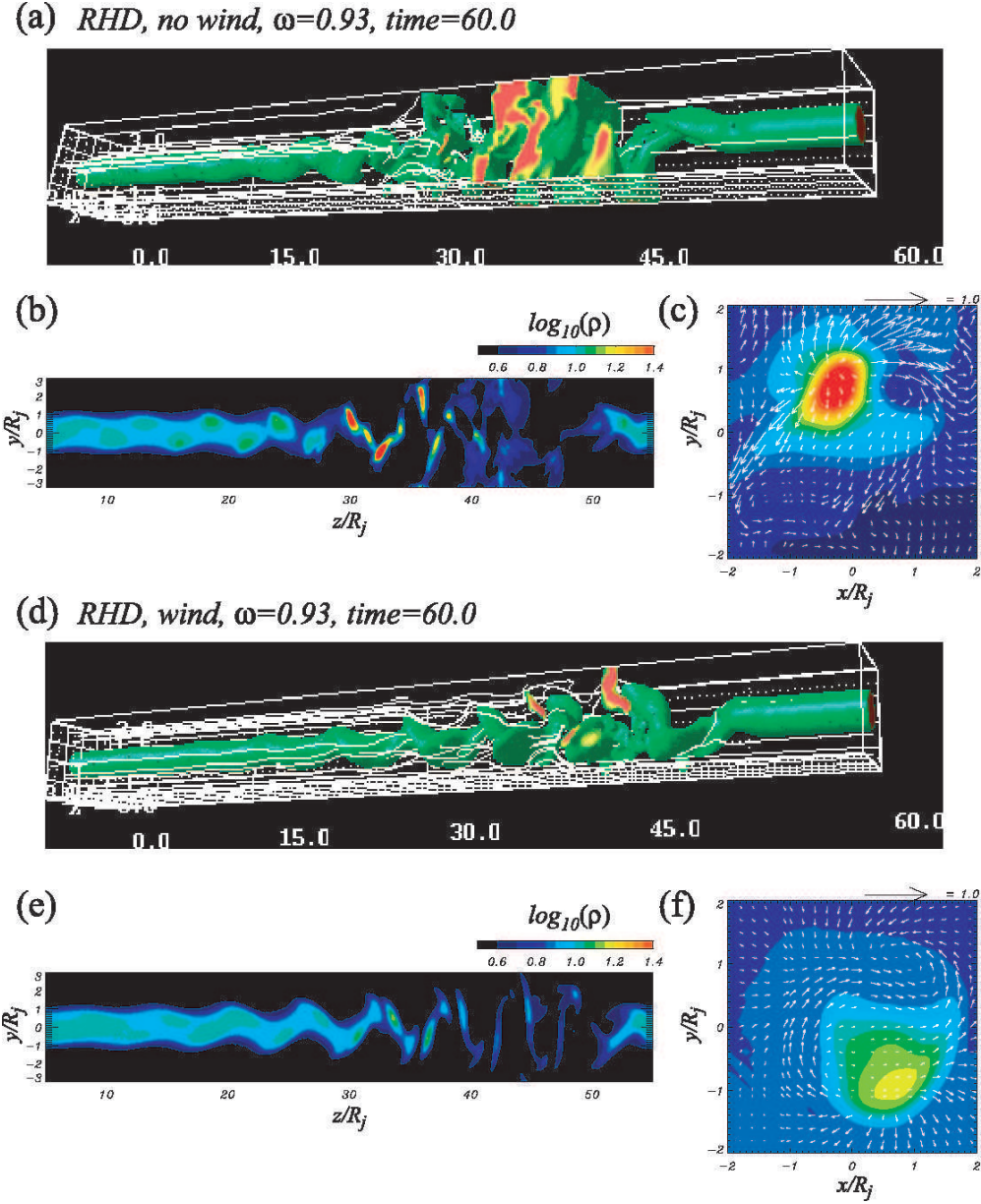}
\caption{Three-dimensional
isovolume image ({\it panels a, d}) and two-dimensional axial ({\it
panels b, e}) and transverse ({\it panels c, f})
slices made at $z=30 R_{j}$ and simulation time $t=60$ for the
weakly magnetized
cases with precession frequency $\omega = 0.93$.  Panels a, b \& c
are for no wind (RHDBn) and panels d, e \& f are with a wind
(RHDBw). The colors show the logarithm of density, white lines
indicate magnetic field lines ({\it a, d}), and arrows depict
transverse velocities. \label{f1}}
\end{figure}

\begin{figure}[hp!]
\epsscale{0.8}
\plotone{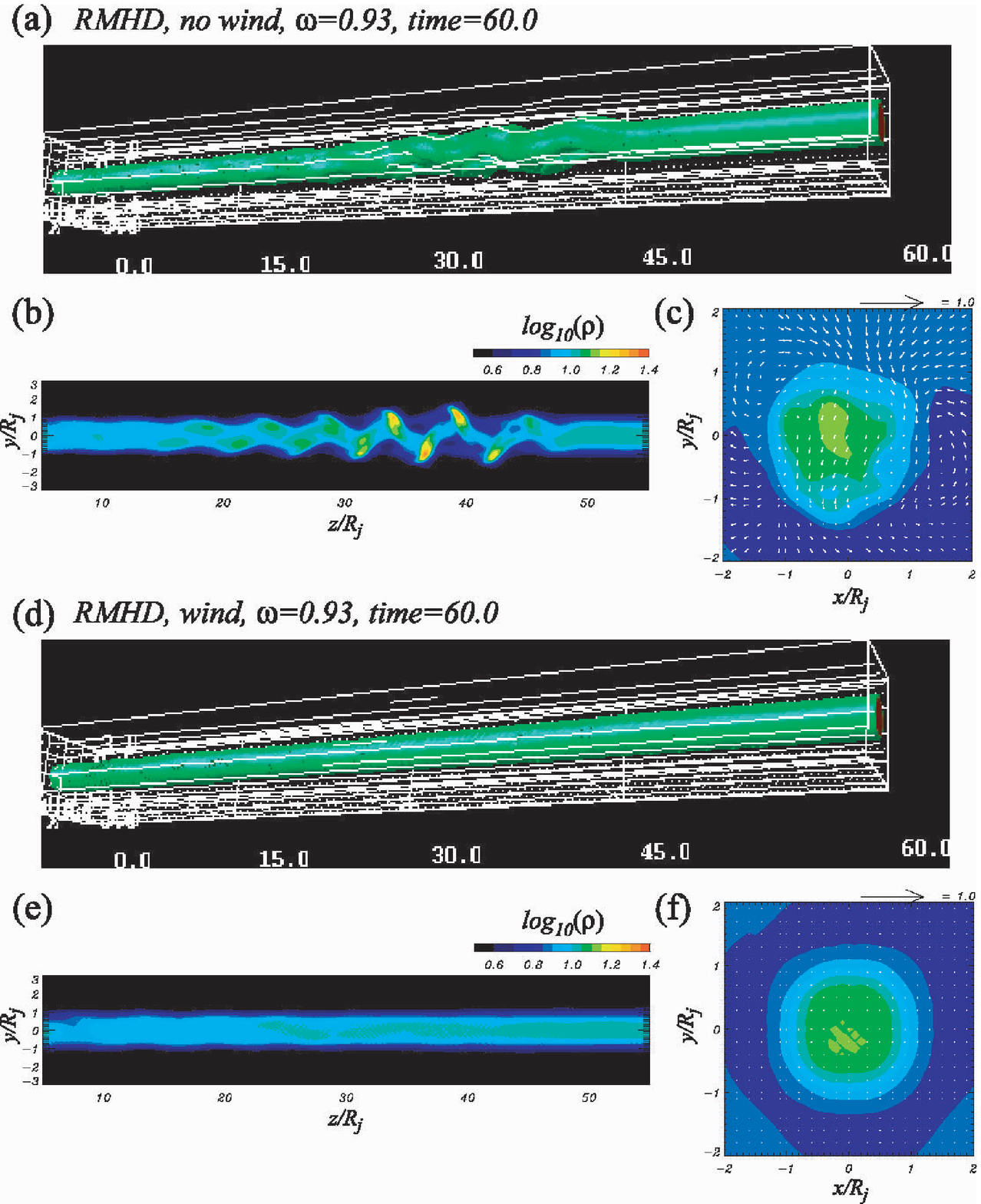}
\caption{Three-dimensional
isovolume image ({\it panels a, d}) and two-dimensional axial ({\it
panels b, e}) and transverse ({\it panels c, f})
slices made at $z=30 R_{j}$ and simulation time $t=60$ for the strongly
magnetized
cases with precession frequency $\omega = 0.93$.  Panels a, b \& c
are for no wind (RMHDBn) and panels d, e \& f are with a wind
(RMHDBw). The colors show the logarithm of density, white lines
indicate magnetic field lines ({\it a, d})and arrows depicts
transverse velocities. \label{f2}}
\end{figure}

Figure $\ref{f1}$ illustrates the difference in jet structure for weak
magnetization with no wind and with an external wind.  Specifically we
show results from the intermediate precession frequency, $\omega 2
\equiv \omega R_{j}/u_{j} = 0.93$, (cases RHDBn
and RHDBw) at simulation time $t=60$. Precession at the jet inflow
plane excites the helical KH mode which is advected down the jet and
grows. The isovolume image shows that beyond $z \sim 30 R_{j}$ with no
wind jet flow is disrupted, and the magnetic field is strongly bent
and distorted. Transverse 2D slices through the jet axis (panels 1b \&
1c) show a smaller density (pressure) fluctuation associated with the
leading edge of the helix in the presence of the external
wind. Transverse 2D slices perpendicular to the jet axis at $z=30
R_{j}$ (panels 1c \& 1f), suggest a less distorted jet in the presence
of the external wind. Transverse velocities shown by the arrows
indicate a circulation around the jet associated with the helical twist that
is much more regular in the presence of the external wind.

Figure $\ref{f2}$ illustrates jet structure for the strongly
magnetized cases with no wind and with an external wind.  As in
Figure 1 we show results for the intermediate precession frequency,
$\omega 2$, (cases RMHDBn and RMHDBw) at simulation time $t=60$. In
the no wind case the helical KH mode grows but more slowly than in
the weakly magnetized cases shown in Figure 1 and does not disrupt
the jet inside $z \sim 40 R_{j}$. A weakly twisted helical flow and
magnetic structure develops. The transverse slice at $z=30 R_{j}$
(panel 2c) indicates weak interaction between the jet and the
external medium at this distance. Some circular circulation is seen.
In the strongly magnetized case with the external wind (RMHDBw), the
helical KH mode is damped and can just barely be seen out to $z = 35
R_{j}$ in the transverse axial slice (panel 2e). The transverse
slice at $z=30 R_{j}$ indicates negligible interaction between jet
and external medium and no circular circulation (panel 2f).

It is immediately clear from Figure 1 that presence of the
wind provides a stabilizing influence in the weakly magnetized
case. Further comparison with Figure 2 shows that the presence of
a strong magnetic field provides a stabilizing influence and complete
stabilization in the presence of the strongly magnetized wind.

To investigate simulation results quantitatively, we take
one-dimensional cuts through the computational box parallel to the
z-axis at radial distances $x/R_{j}=0.2$, $0.5$, and $0.8$ on the
transverse x-axis. The results for weakly magnetized cases
with/without the external wind are shown in Figures 3, 4, \& 5 for
precession frequencies, $\omega R_{j}/u_{j} = 0.4,~0.93,~\&~2.69$
respectively.  The results for the strongly magnetized cases are shown
in Figures 6 \& 7 for precession frequencies $\omega R_{j}/u_{j} =
0.93,~\&~2.69$ respectively. In the figures, $u_{x}$ and $u_{y}$
velocity components correspond to radial $u_{r}$ and azimuthal
$u_{\phi}$ velocity components in cylindrical geometry.

In the weakly magnetized cases oscillation from the growing helical KH
mode is seen in all cases. From the plots of radial and transverse
velocities, the dominant wavelengths of oscillation are $\lambda/R_{j}
\sim 13,~6,~\&~2$ for low, intermediate, and high frequency precession
with no wind and with a measurable lengthening to $\lambda/R_{j} \sim 14$
for low frequency precession with the wind.  For the high-frequency
case without the external wind, the azimuthal velocity component
suggests a beat pattern with wavelength, $\lambda^{n}_{beat}(\omega
3) \lesssim 20 R_{j}$.  From the axial velocities near the jet axis,
we see that jet flow is disrupted at $z \sim 25 R_{j}$, $z \sim 32
R_{j}$, and possibly for $z \gtrsim 50 R_{j}$ for the low,
intermediate and high precession frequency no wind cases
respectively. Jet flow is disrupted at $z \sim 43 R_{j}$, $z \sim 35
R_{j}$ and possibly for $z \gtrsim 50 R_{j}$ for the low, intermediate
and high frequency precession wind cases respectively.  Thus, the
external wind reduces the growth of KH instability and delays the
onset of flow disruption. The large dips in the axial velocity near the jet
surface are caused by sideways motion of the jet surface. Dips in the axial
velocity occur more deeply inside the jet surface for lower frequency
precession.  The presence of the external wind reduces these effects
somewhat.

Quantitative simulation results for the strongly magnetized cases
with/without the external wind are shown in Figures 6 and 7. In the
absence of the external wind (RMHDBn and RMHDCn) growing oscillation
from helical KH instability is seen. However, spatial growth is much
slower than for the comparable weakly magnetized cases. Comparison
with the weakly-magnetized cases (RHDBn and RHDCn), shows that the
magnetic field reduces the growth rate of KH instabilities such that
any disruption of collimated flow will lie at $z \gg 40 R_{j}$. In
the presence of the external wind (RMHDBw and RMHDCw) the initial
oscillation is damped. Damping is more rapid for the intermediate
frequency perturbation than for the high frequency perturbation.
Plots of radial and transverse velocities, and the radial magnetic
field component shown in one-dimensional cuts in Figures 6 and 7
show that the dominant wavelengths of oscillation are $\lambda/R_{j}
\sim 5~\&~2$ for the intermediate and high frequency perturbations,
respectively with or without the external wind. There is a possible
beat pattern in the azimuthal velocity component with
$\lambda^{mn}_{beat} (\omega 2) \sim 10~R_{j}$ for $z < 20~R_{j}$ in
the no wind high frequency case. The presence of the wind results in
more readily seen beat patterns accompanying the damped
oscillations. The beat pattern is best seen in the radial magnetic
field amplitude (Fig.\ 6f and Fig.\ 7f).  At the intermediate
perturbation frequency the beat wavelength is $\lambda^{mw}_{beat}
(\omega 2) \gtrsim 20~R_{j}$. In the high-frequency case a clear
beat pattern with $\lambda^{mw}_{beat}(\omega 3) \sim 10~R_{j}$ is
seen at $z < 33~R_{j}$ but disappears at larger $z$.

\begin{figure}[hp!]
\epsscale{0.75}
\plotone{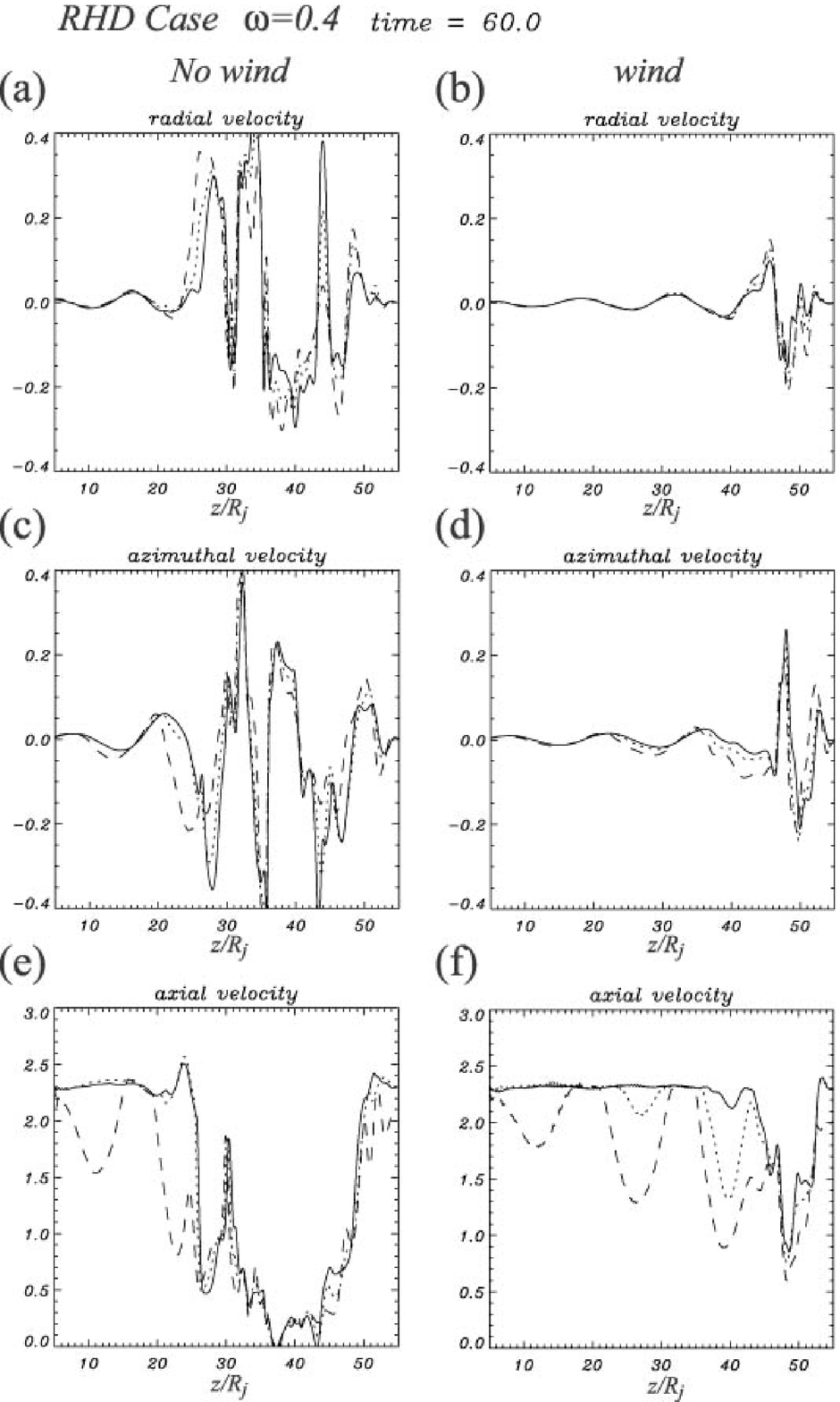}
\caption{Radial velocity ($u_{x}$), azimuthal velocity
($u_{y}$), and axial velocity ($\gamma u_{z}$) along the one-dimensional cuts
parallel to the jet axis located at $x/R_{j}= 0.2$(solid line),
$0.5$(dotted line) and $0.8$(dashed line) for low-frequency precession
of a weakly magnetized jet for no wind (RHDAn)(left panels)
and with an external wind (RHDAw)(right panels). \label{f3}}
\end{figure}

\begin{figure}[hp!]
\epsscale{0.75}
\plotone{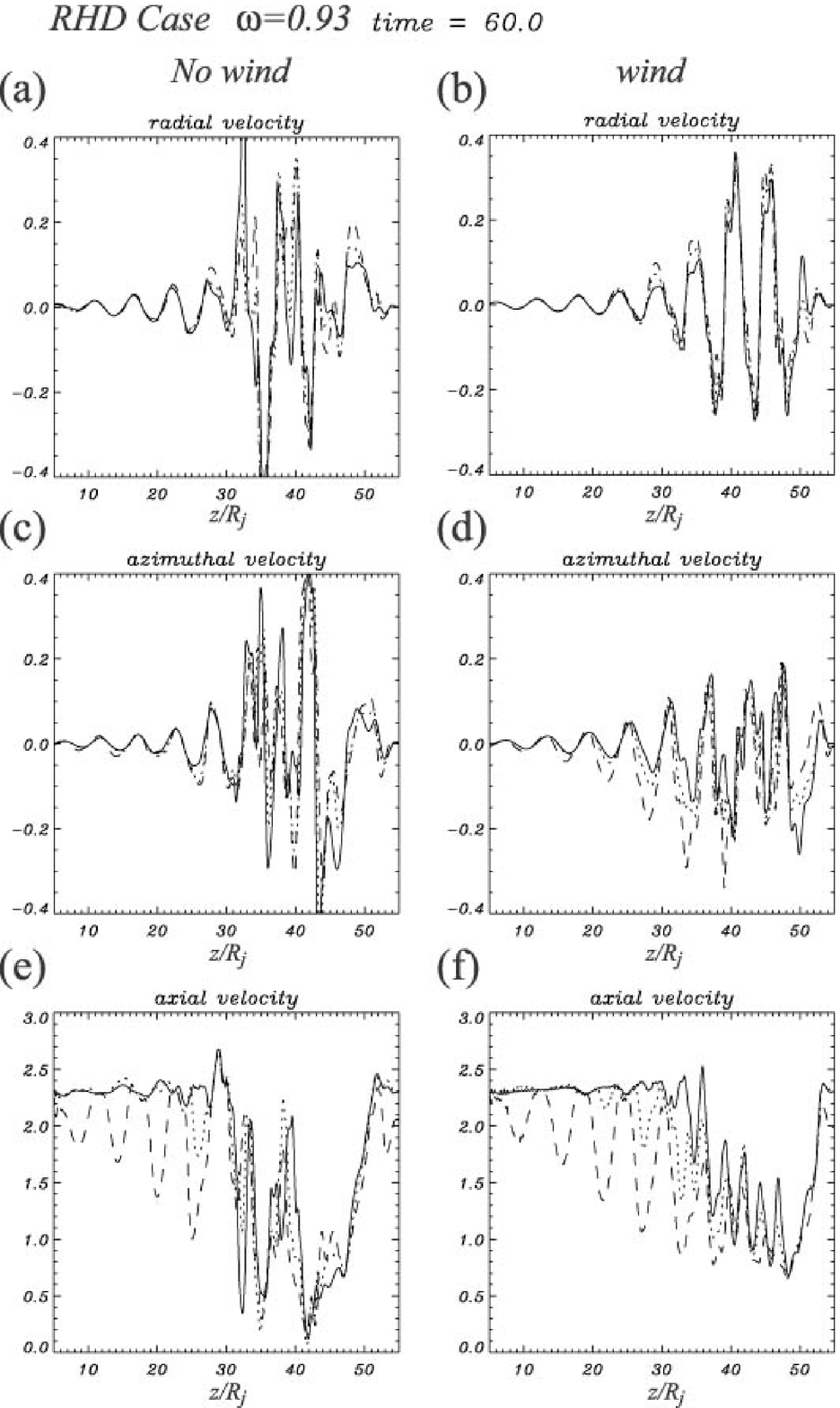}
\caption{Radial velocity ($u_{x}$), azimuthal velocity
($u_{y}$), and axial velocity ($\gamma u_{z}$) along the one-dimensional cuts
parallel to the jet axis located at $x/R_{j}= 0.2$(solid line),
$0.5$(dotted line) and $0.8$(dashed line) for intermediate frequency
precession of a weakly magnetized jet for no wind
(RHDBn)(left panels) and with an external wind (RHDBw)(right panels).
\label{f4}}
\end{figure}

\begin{figure}[hp!]
\epsscale{0.75}
\plotone{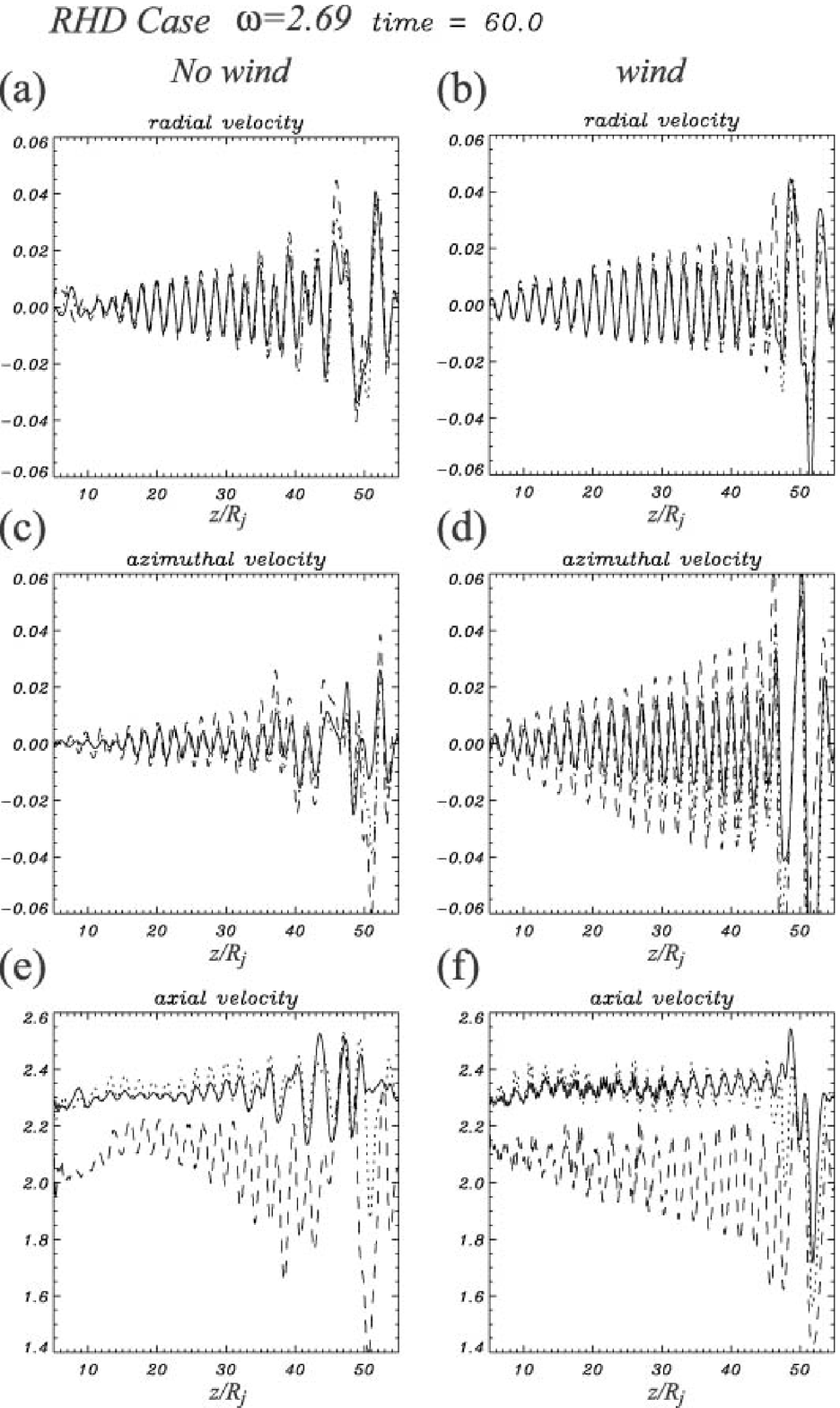}
\caption{Radial velocity ($u_{x}$), azimuthal velocity
($u_{y}$), and axial velocity ($\gamma u_{z}$) along the one-dimensional cuts
parallel to the jet axis located at $x/R_{j}= 0.2$(solid line),
$0.5$(dotted line) and $0.8$(dashed line) for high frequency
precession of a weakly magnetized jet for no wind (RHDCn)(left panels)
and with an external wind (RHDCw)(right panels). \label{f5}}
\end{figure}

\begin{figure}[hp!]
\epsscale{0.75}
\plotone{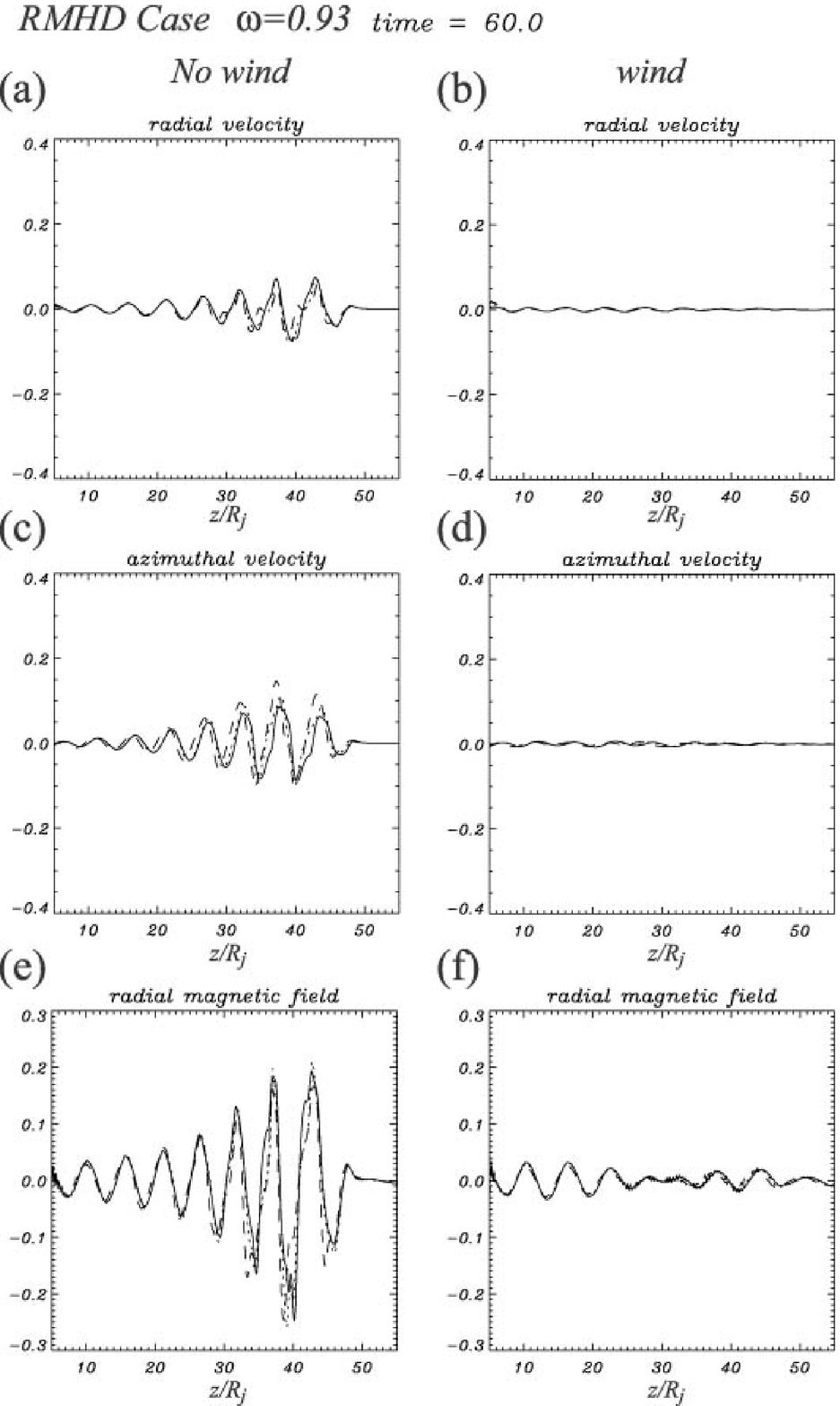}
\caption{Radial velocity ($u_{x}$), azimuthal velocity
($u_{y}$), and radial magnetic field ($B_{x}$) along the
one-dimensional cuts parallel to the jet axis located at $x/R_{j}=
0.2$(solid line), $0.5$(dotted line) and $0.8$(dashed line) for
intermediate frequency precession of a strongly magnetized jet for no
wind (RMHDBn)(left panels) and with a strongly magnetized external wind
(RMHDBw)(right panels). \label{f6}}
\end{figure}

\begin{figure}[hp!]
\epsscale{0.75}
\plotone{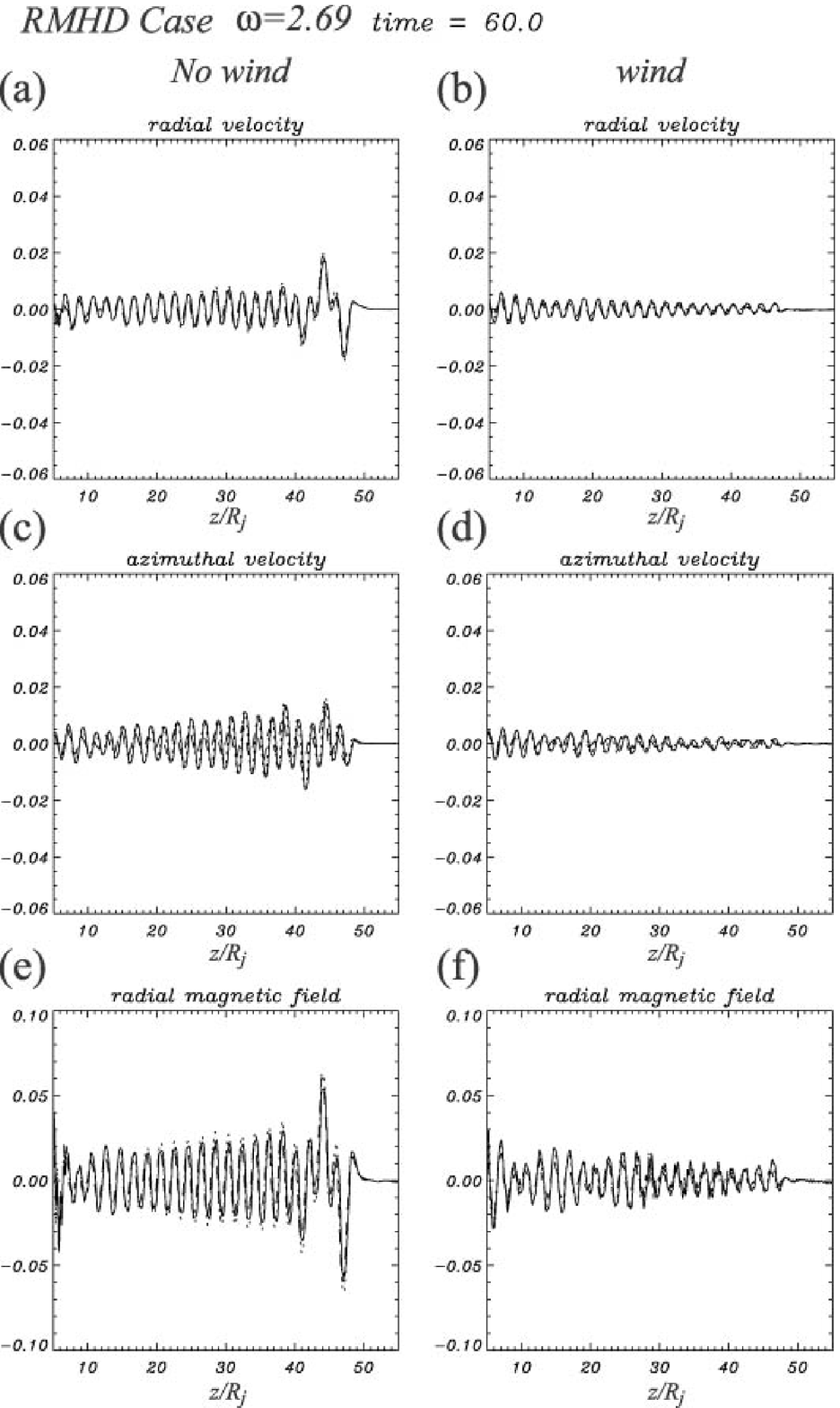}
\caption{Radial velocity ($u_{x}$), azimuthal velocity
($u_{y}$), and radial magnetic field ($B_{x}$) along the
one-dimensional cuts parallel to the jet axis located at $x/R_{j}=
0.2$(solid line), $0.5$(dotted line) and $0.8$(dashed line) for high
frequency precession of a strongly magnetized jet for no wind (RMHDCn)
(left panels) and with a strongly magnetized external wind (RMHDCw)(right
panels). \label{f7}}
\end{figure}

In the next section we will compare these simulation results with
theoretical predictions for growth and damping of the helical wave
mode excited by the inlet perturbations in our numerical simulations.

\newpage

\section{Theory and Analysis}

\subsection{Dispersion Relation}

Stability of a jet spine-sheath or jet-wind configuration, where the
sheath or wind is much broader than the spine or jet, can be
accomplished by modeling the jet/spine as a cylinder of radius R
embedded in an infinite wind/sheath.

Formally, the assumption of an infinite sheath means that the analysis
could be performed in the reference frame of the sheath
and numerical simulations could be performed in the reference frame
of the sheath with results transformed to the source/observer reference
frame. However, it is not much more difficult to derive a dispersion
relation and obtain analytical expressions in the source/observer frame
and analytical solutions to the dispersion relation in the
source/observer frame take on simple
physically revealing forms (Hardee 2007). Additionally, this approach
lends itself to modeling the propagation and appearance of jet structures
viewed in the source/observer frame, e.g., helical structures in the 3C\,120
jet (Hardee et al.\ 2005).

A dispersion relation
describing the growth or damping of the normal wave modes associated
with this system can be derived if uniform conditions are assumed
within the jet/spine, e.g., having a uniform proper density, $\rho
_{j}$, a uniform axial magnetic field, $B_{j}=B_{j,z}$, and a
uniform velocity, $\mathbf{u}_{j}=u_{j,z}$, and if uniform
conditions are assumed within the external sheath/wind, e.g., having
a uniform proper density, $ \rho _{e}$, a uniform axial magnetic
field, $B_{e}=B_{e,z}$, and a uniform velocity
$\mathbf{u}_{e}=u_{e,z}$. Here the jet/spine is established in
static total pressure balance with the external wind/sheath where
the total static uniform pressure is $P_{e}^{\ast }\equiv
P_{e}+B_{e}^{2}/8\pi =P_{j}^{\ast }\equiv P_{j}+B_{j}^{2}/8\pi$.

The dispersion relation is obtained from the linearized ideal RMHD and
Maxwell equations, where the density, velocity, pressure and magnetic
field are written as $\rho =\rho _{0}+\rho _{1}$,
$\mathbf{v}=\mathbf{u}+\mathbf{v}_{1}$ (we use $\mathbf{v}_{0}\equiv
\mathbf{u}$ for notational reasons), $P=P_{0}+P_{1}$, and $\mathbf{B}=
\mathbf{B}_{0}+\mathbf{B}_{1}$, where subscript 1 refers to a
perturbation to the equilibrium quantity with subscript
0. Additionally, the Lorentz factor $\gamma ^{2}=(\gamma _{0}+\gamma
_{1})^{2}\simeq \gamma _{0}^{2}+2\gamma _{0}^{4}\mathbf{u\cdot
v}_{1}/c^{2}$ where $\gamma _{1}=\gamma _{0}^{3}\mathbf{u\cdot
v}_{1}/c^{2}$.  It is assumed that the initial equilibrium system
satisfies the zero order equations. Details of the derivation for the
fully relativistic case can be found in Hardee (2007).

In cylindrical geometry a random perturbation $\rho _{1}$,
$\mathbf{v}_{1}$ $ \mathbf{B}_{1}$ and $P_{1}$ can be considered to
consist of Fourier components of the form
\begin{equation}
f_{1}(r,\phi ,z,t)=f_{1}(r)\exp [i(kz\pm n\phi -\omega t)]  \label{3}
\end{equation}
where the zero order flow is along the z-axis, and r is in the radial
direction with the jet/spine bounded by $r=R$. In cylindrical geometry
$n$ is an integer azimuthal wavenumber, for $n>0$ waves propagate at
an angle to the flow direction, and $+n$ and $-n$ give wave
propagation in the clockwise and counter-clockwise sense,
respectively, when viewed in the flow direction. In equation (1)
$n=0$, $1$, $2$, $3$, $4$, etc. correspond to pinching, helical,
elliptical, triangular, rectangular, etc. normal mode distortions of
the jet, respectively. Propagation and growth or damping of the
Fourier components can be described by a dispersion relation of the
form
\begin{equation}
\frac{\beta _{j}}{\chi _{j}}\frac{J_{n}^{^{\prime }}(\beta _{j}R)}{
J_{n}(\beta _{j}R)}=\frac{\beta _{e}}{\chi _{e}}\frac{H_{n}^{(1)^{\prime
}}(\beta _{e}R)}{H_{n}^{(1)}(\beta _{e}R)}~.  \label{4}
\end{equation}
In the dispersion relation $J_{n}$ and $H_{n}^{(1)}$ are Bessel and
Hankel functions, and the primes denote derivatives of the Bessel and
Hankel functions with respect to their arguments. In equation (4)
\begin{equation}
\eqnum{5a}
\chi _{j}\equiv \gamma _{j}^{2}\gamma _{Aj}^{2}W_{j}\left( \varpi
_{j}^{2}-\kappa _{j}^{2}v_{Aj}^{2}\right)~,  \label{5a}
\end{equation}
\begin{equation}
\eqnum{5b}
\chi _{e}\equiv \gamma _{e}^{2}\gamma _{Ae}^{2}W_{e}\left( \varpi
_{e}^{2}-\kappa _{e}^{2}v_{Ae}^{2}\right)~,  \label{5b}
\end{equation}
\setcounter{equation}{5}
and
\begin{equation}
\eqnum{6a}
\beta _{j}^{2}\equiv \left[ \frac{\gamma _{j}^{2}\left( \varpi
_{j}^{2}-\kappa _{j}^{2}a_{j}^{2}\right) \left( \varpi _{j}^{2}-\kappa
_{j}^{2}v_{Aj}^{2}\right) }{v_{msj}^{2} \varpi _{j}^{2}-\kappa
_{j}^{2}v_{Aj}^{2}a_{j}^{2}}\right]~,
\label{6a}
\end{equation}
\begin{equation}
\eqnum{6b}
\beta _{e}^{2}\equiv \left[ \frac{\gamma _{e}^{2}\left( \varpi
_{ex}^{2}-\kappa _{e}^{2}a_{e}^{2}\right) \left( \varpi
_{e}^{2}-\kappa _{e}^{2}v_{Ae}^{2}\right) }{v_{mse}^{2} \varpi
_{e}^{2}-\kappa _{e}^{2}v_{Ae}^{2}a_{e}^{2}}\right]~.
\label{6b}
\end{equation}
\setcounter{equation}{6}
In equations (5a \& 5b) and equations (6a \& 6b) $\varpi
_{j,e}^{2}\equiv \left( \omega -ku_{j,e}\right) ^{2}$ and $\kappa
_{j,e}^{2}\equiv \left( k-\omega u_{j,e}/c^{2}\right) ^{2}$, $\gamma
_{j,e}\equiv (1-u_{j,e}^{2}/c^{2})^{-1/2} $ is the flow Lorentz
factor, $\gamma _{Aj,e}\equiv (1-v_{Aj,e}^{2}/c^{2})^{-1/2}$ is the
Alfv\'{e}n Lorentz factor, $W\equiv \rho +\left[ \Gamma /\left(
\Gamma -1\right) \right] P/c^{2} $ is the enthalpy, $a$ is the sound
speed, $v_{A}$ is the Alfv\'{e}n wave speed, and $v_{ms}$ is a
magnetosonic speed. The sound speed is defined by
$$
a\equiv \left[ \frac{\Gamma P}{\rho +\left( \frac{\Gamma }{\Gamma -1}\right)
P/c^{2}}\right] ^{1/2}~,
$$
where $4/3\leq \Gamma \leq 5/3$ is the adiabatic index. The
Alfv\'{e}n wave speed defined by
$$
v_{A}\equiv \left[ \frac{V_{A}^{2}}{1+V_{A}^{2}/c^{2}}\right] ^{1/2}
$$
where $V_{A}^{2}\equiv B_{0}^{2}/(4\pi W_{0})$ is equivalent to
equation (2). A magnetosonic speed
corresponding to the fast magnetosonic speed for propagation
perpendicular to the magnetic field (e.g., Vlahakis \& K\"onigl 2003)
is defined by
$$
v_{ms} \equiv \left[a^2 + v_A^2 - a^2v_A^2/c^2\right]^{1/2} =
\left[a^2/\gamma_A^2 + v_A^2\right]^{1/2}~.
$$

Each normal mode $n$ contains a single \textit{fundamental wave}
($\omega \rightarrow 0$, $k\rightarrow 0$, $\omega /k>0$) and
multiple \textit{body wave}($\omega \rightarrow 0$, $k>0$, $\omega
/k\rightarrow 0$) solutions that satisfy the dispersion relation. In
the numerical simulations the jet/spine was precessed in a manner
designed to trigger the $n = 1$ fundamental helical mode. Previous
theoretical and simulation work has shown a sufficiently close
coupling between fundamental and body modes that the simulation may
excite the first body mode as well. In what follows here we consider
the $n = 1$ helical fundamental and first body mode solutions to the
dispersion relation as being relevant to the numerical simulations
performed here.

\subsection{The Helical Mode}

In the low frequency limit the helical fundamental mode has an analytic
wave solution given by
\begin{equation}
\frac{\omega }{k}=\frac{\left[ \eta u_{j}+u_{e}\right] \pm i\eta
^{1/2}\left[ \left( u_{j}-u_{e}\right) ^{2}-V_{As}^{2}/\gamma
_{j}^{2}\gamma _{e}^{2} \right] ^{1/2}}{(1+V_{Ae}^{2}/\gamma
_{e}^{2}c^{2})+\eta (1+V_{Aj}^{2}/\gamma _{j}^{2}c^{2})} \label{7}
\end{equation}
where $\eta \equiv \gamma _{j}^{2}W_{j}\left/ \gamma
_{e}^{2}W_{e}\right. $ and a ``surface'' Alfv\'{e}n speed is given by
\begin{equation}
V_{As}^{2}\equiv \left( \gamma _{Aj}^{2}W_{j}+\gamma
_{Ae}^{2}W_{e}\right) \frac{B_{j}^{2}+B_{e}^{2}}{4\pi W_{j}W_{e}}~.
\label{8}
\end{equation}
In equation (8) note that the Alfv\'{e}n Lorentz factor can be written
as $\gamma _{Aj,e}^{2}=1+V_{Aj,e}^{2}/c^{2}$. The jet is predicted to
be stable to the helical fundamental mode when
\begin{equation}
\left( u_{j}-u_{e}\right)^{2}-V_{As}^{2}/\gamma _{j}^{2}\gamma
_{e}^{2} < 0~.
\label{9}
\end{equation}
Thus, as might be anticipated, the growth rate is directly
related to the difference between the magnitude of a ``shear'' speed,
$(u_j - u_e)^2$, and a ``surface'' Alfv\'{e}n speed. Note that the
``surface'' Alfv\'{e}n speed can be greater than the speed of light
and is not a physical wave speed. The growth rate is also reduced by
the spine Lorentz factor through $\eta$ in the denominator of eq.\
(7).  Finally, the real part of eq.\ (7) directly provides an estimate
of the increase in helical pattern speed resulting from the external
sheath flow, and this increase when combined with a decrease in the
temporal growth rate implies an increase in the spatial growth
length.

In the low frequency limit the real part of the first helical
\textit{body wave} solution has an analytic solution given
approximately by
\begin{equation}
kR\approx k^{\min }R\equiv \frac{5}{4} \pi \left[ \frac{
v_{msj}^2u_{j}^{2}-v_{Aj}^2a_{j}^{2}}{\gamma
_{j}^{2}(u_{j}^{2}-a_{j}^{2})(u_{j}^{2}-v_{Aj}^{2})}\right]^{1/2}~.
 \label{10}
\end{equation}
In this low frequency limit the body wave solution exists only when
$k^{\min }R$ has a positive real part.  This requires that
\begin{equation}
 \left[ \frac{ v_{msj}^2u_{j}^{2}-v_{Aj}^2a_{j}^{2}}{\gamma
 _{j}^{2}(u_{j}^{2}-a_{j}^{2})(u_{j}^{2}-v_{Aj}^{2})}\right] >0~.
\label{11}
\end{equation}
Thus, the first helical body mode exists when the jet is supersonic
and super-Alfv\'{e}nic, i.e., $u_{j}^{2}-a_{j}^{2}>0$ and
$u_{j}^{2}-v_{Aj}^{2}>0$, or in a limited velocity range given
approximately by $a_{j}^{2}>u_{j}^{2}>[\gamma _{sj}^{2}/(1+\gamma
_{sj}^{2})]a_{j}^{2}$ when $v_{Aj}^{2}\approx a_{j}^{2}$, where
$\gamma _{sj} \equiv (1-a_{j}^2/c^2)^{-1/2}$ is a sonic Lorentz
factor.

For a supermagnetosonic jet, the helical fundamental and first body
modes can have a distinct maximum in the growth rate at some
resonant frequency.  The resonance condition can be evaluated
analytically in either the fluid limit where $a \gg V_{A}$ or in the
magnetic limit where $V_{A} \gg a$. Note that in the magnetic limit,
magnetic pressure balance implies that $B_j = B_e$. In these cases a
necessary condition for resonance is that
\begin{equation}
\frac{u_{j}-u_{e}}{1-u_{j}u_{e}/c^{2}}>\frac{v_{wj}+v_{we}}{
1+v_{wj}v_{we}/c^{2}}~,  \label{12}
\end{equation}
where $v_{wj}\equiv \left( a_{j},v_{Aj}\right) $ and $v_{we}\equiv
\left( a_{e},v_{Ae}\right) $ in the fluid or magnetic limits,
respectively. This necessary condition for resonance
indicates that we are supersonic or super-Alfv\'enic when the shear
speed exceeds a physical ``surface'' wave speed.  When this condition
is satisfied it can be shown that the wave speed at resonance is
\begin{equation}
v_{w} \approx v_{w}^{\ast }\equiv \frac{\gamma _{j}(\gamma
_{we}v_{we})u_{j}+\gamma _{e}(\gamma _{wj}v_{wj})u_{e}}{\gamma
_{j}(\gamma _{we}v_{we})+\gamma _{e}(\gamma _{wj}v_{wj})} \label{13}
\end{equation}
where $\gamma _{w}\equiv (1-v_{w}^{2}/c^{2})^{-1/2}$ is the sonic or
Alfv\'{e}n Lorentz factor accompanying $v_{wj}\equiv \left(
a_{j},v_{Aj}\right) $ and $v_{we}\equiv \left( a_{e},v_{Ae}\right) $
in the fluid or magnetic limits, respectively.  The resonant wave
speed and maximum growth rate occur at a frequency given by
\begin{equation}
\omega R/v_{we} \approx \omega_{m}^{\ast}R/v_{we} \equiv
\frac{3\pi/4+m\pi }{ \left[\left(1 - u_{e}/v^{\ast}_w\right)^2 -
\left(v_{we}/v^{\ast}_w - u_e v_{we}/c^2\right)^2\right]^{1/2}}~.
\label{14}
\end{equation}
In equation (14) $m=0,1$ specifies the fundamental and first body
modes, respectively.  A resonant wavelength is given by $\lambda
\approx \lambda _{m}^{\ast }\equiv 2\pi v_{w}^{\ast }/\omega_{m}^{\ast
}$ and can be calculated from
\begin{equation}
\lambda _{m}^{\ast }\equiv\frac{2\pi}{3\pi
  /4+m\pi}\left(\frac{\gamma_e}{v_{we}}\right)\left\{\left(v_{w}^{\ast} -
    u_e\right)^2 - \left[v_{we} -
    (v_{we}u_e/c^2)v_{w}^{\ast}\right]^2\right\}^{1/2} R ~. \label{15}
\end{equation}

The resonant frequency is found to be largely a function of the sound
and Alfv\'en wave speeds in the sheath and the shear speed, $u_j -
u_e$ (Hardee 2007).  The resonant frequency increases as the sound and
Alfv\'en wave speeds increase and as the shear speed declines. In the
limit
$$
\frac{u_{j}-u_{e}}{1-u_{j}u_{e}/c^{2}}\longrightarrow \frac{v_{wj}+v_{we}}{
1+v_{wj}v_{we}/c^{2}}~,
$$
the resonant frequency $\omega _{m}^{\ast }R/v_{we}\rightarrow
\infty $.  In general, When the sound or Alfv\'{e}n wave speed
increases relative to the jet speed there is an increase in the
growth rate at the higher resonant frequency accompanying an
increase in the sound or Alfv\'{e}n wave speed relative to the jet
speed.  On the other hand, the growth rate at resonance decreases as
the shear speed, $u_j - u_e$, declines.  This decline in the growth
rate is also indicated by equation (8) which applies to fundamental
mode frequencies up to an order of magnitude below resonance. As the
resonant frequency increases equation (8) applies to increasingly
higher fundamental mode frequencies.

Numerical solution of the dispersion relation is necessary to obtain
accurate values for growth or damping rates as fundamental mode
frequencies approach and exceed the resonant frequency. In general,
the behavior of growth or damping associated with the first body
mode must be obtained by numerical solution of the dispersion
relation at all frequencies.  In the high frequency limit the real
part of the fundamental and first body mode solutions to the
dispersion relation tend towards the analytic limiting form
\begin{equation}
\frac{\omega }{k}\approx \frac{u_{j}\pm v_{wj}}{1\pm
v_{wj}u_{j}/c^{2}}~.  \label{16}
\end{equation}
which describes sound waves $v_{wj}=a_{j}$ or Alfv\'{e}n waves $
v_{wj}=v_{Aj}$ propagating with and against the jet flow inside the
jet.
Note that at high frequencies waves propagate in the spine
  fluid with speeds that are independent of the surrounding sheath
  and are decoupled from the spine sheath boundary.

\subsection{Numerical Solution to the Dispersion Relation}

In general, equations (7) and (10) provide initial estimates at low
frequencies to the helical fundamental and first body mode solutions
that can then be followed by root finding techniques to higher
frequencies.  The results of numerical solution to the dispersion
relation for the parameters appropriate to the numerical simulations
shown in section 2 are displayed in Figure 8. It should be noted
that not all possible solutions are shown or have necessarily been
found by the root finding technique. In general, in the weakly
magnetized case fundamental (S) mode solutions consist of a growing
(shown) and damped (not shown) solution pair with comparable growth
and damping rates (see eq.\ 7) and first body (B1) mode solutions
consist of a real and growing or damped solution pair. The presence
of the external wind flow leads to reduced growth of the S mode and
weak damping of the B1 mode.
\begin{figure}[hp!]
\epsscale{0.99}
\plotone{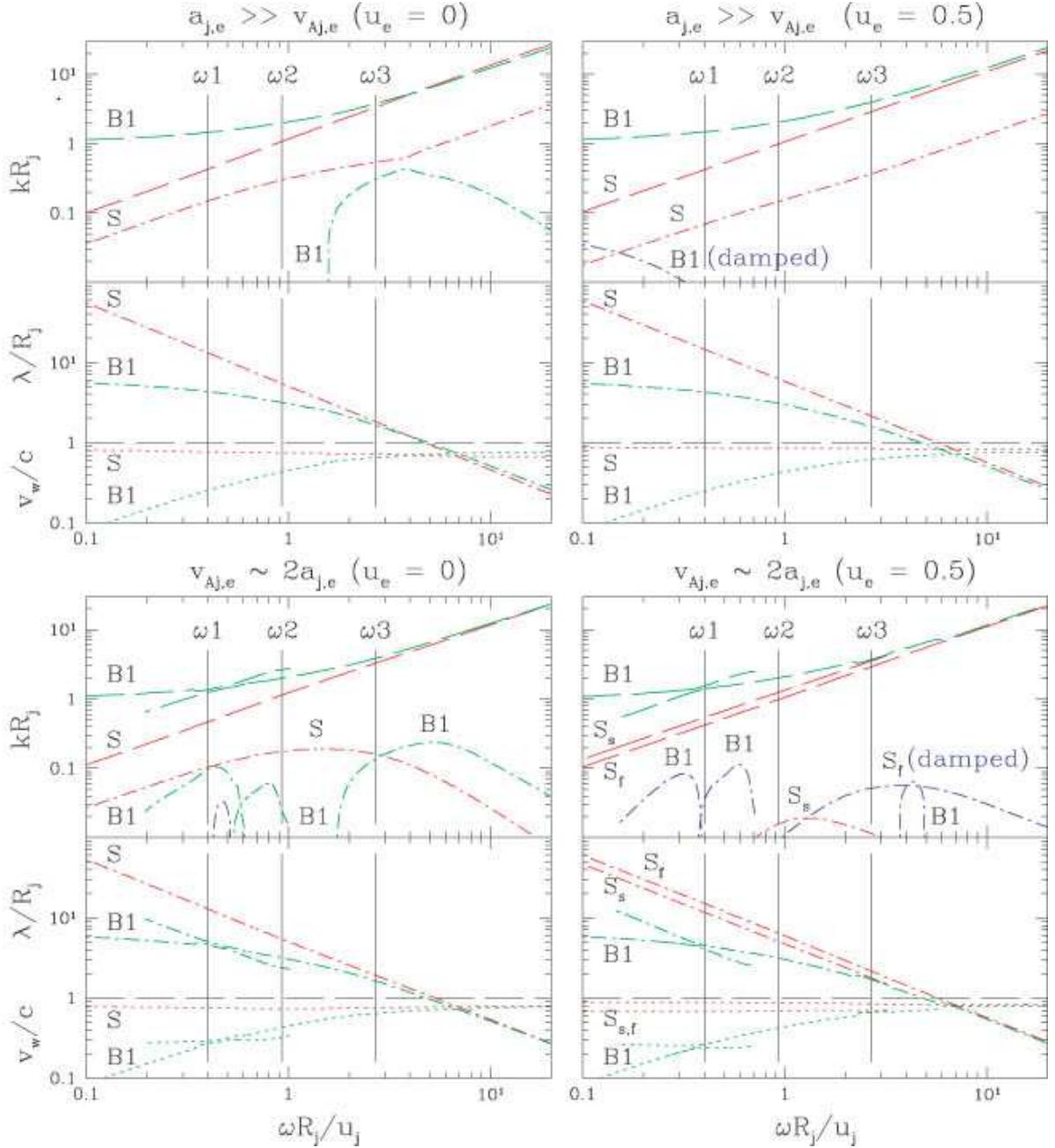}
\caption{Solutions of the dispersion relation for helical fundamental
(red lines) and first body (green lines) modes for weakly magnetized
($a_{j,e} \gg v_{Aj,e}$) and strongly magnetized ($a_{j,e} \sim
v_{Aj,e}$) jet simulations without a surrounding outflow ($u_e = 0$)
and with a surrounding 0.5~c outflow ($u_e = 0.5$).  Dispersion
relation solutions show the real, $k_rR_j$, (dashed lines) and
imaginary, $k_iR_j$, (dash-dot lines) parts of the dimensionless
wavenumber normalized by the jet radius, $R_j$, as a function of the
dimensionless angular frequency, $\omega R_j/ u_j$, normalized by the
jet radius and jet speed, $u_j$. Where the imaginary part of the
wavenumber is shown in blue, the solution is damped.  Immediately
under the panel showing a dispersion relation solution for fundamental
(S) and first body (B1) modes is a panel that shows the wavelength,
$\lambda/R_j$, (dash-dot lines) and wave speed, $v_w/c$, (dotted
lines).  The angular driving frequencies used in the numerical
simulations are indicated by the vertical solid lines.
\label {f8}}
\end{figure}

The solution structure is more complex in the strongly magnetized
case.  Comparable sound and Alfv\'en wave speeds have led to an
increased complexity compared to the weakly magnetized case solutions
or when compared to the analytically predicted results for Alfv\'en
wave speed greatly exceeding the sound speed. In the absence of
magnetized sheath flow S mode solutions again consist of a growing
(shown) and damped (not shown) solution pair.
Now however, we find
multiple growing solutions associated with the B1 mode evident at the
lower frequencies (see the lower left panel in Figure 8). A modest
damping rate accompanies the crossing of the real part of these
multiple body mode solutions.
The complex structure of the first body mode solution shown in
Figure 8 for the strongly magnetized cases has been seen previously in
a non-relativistic stability analysis (see Figure 20 in Hardee et al.\
1995). At the lowest frequencies, $\omega < \omega 1$, the first body
mode consists of a solution pair whose real part can be identified
with the descending and constant real part of the wavenumber. At
frequencies above the crossing point, $\omega > \omega 1$, the rapidly
rising real part of the wavenumber connects to the second body mode at
a frequency $\omega > \omega 2$.

At the higher frequencies the B1 mode is similar to the weakly
magnetized case.  In the presence of magnetized sheath flow, there is
significant difference in growth and damping rates for the S mode
solution pair.  Weak growth is associated with the slower, $S_{s}$,
moving shorter wavelength solution and weak damping is associated with
the faster, $S_{f}$, moving longer wavelength solution.  At the
intermediate frequency, $\omega 2$, and below the growth rate is
larger than the damping rate but at the higher frequencies somewhat above
$\omega 2$ the damping rate is larger than growth rate for the S mode
solution pair.  The presence of magnetized sheath flow leads to
damping of the B1 mode at the lower frequencies where in the absence
of magnetized sheath flow there was modest growth and a modest high
frequency damping rate maximum is seen where B1 intersects $S_{s}$.

The wavelengths and growth/damping lengths normalized to
the jet radius for the precession frequencies $\omega 1$, $\omega 2$
and $\omega 3$ used in the weakly magnetized simulations are given in
Table 2.
\placetable{table 2}
Weak damping ($wd$) indicates a damping length longer than
the grid length in the numerical simulations. No entry in the
growth/damping length column indicates a purely real solution.
Wavelengths observed in the simulations are in excellent agreement
with the theoretically predicted wavelengths.

When no external wind is present the dispersion relation solutions
show growth of the fundamental (S) mode at precession frequencies
$\omega 1$ and $\omega 2$ and growth of the body mode at higher
frequencies.  In the weakly magnetized case coupling between
fundamental and first body (B1) modes is indicated by the solution
structure just above precession frequency, $\omega 3$, where the
real and imaginary parts of the S and B1 mode solutions are
comparable.  Thus, we expect to see an indication of interaction in
the $\omega 3$ weakly magnetized simulation between the S and B1
modes with a beat wavelength of $\lambda_{beat} = 16.8~R_{j}$. This
agrees with the observed $\lambda^{n}_{beat}(\omega 3) \lesssim
20~R_{j}$ in the simulation.

External wind flow in the weakly magnetized case leads to weak damping
of the B1 mode and reduces the growth rate and increases the growth
length, $\ell \equiv k_i^{-1}$, of the S mode by about a factor of two
at all frequencies. However, the numerical simulation for the high
frequency precession of a weakly magnetized jet (RHDC), see
Figure 5, indicates a larger perturbation growth for the wind case than
for the no wind case where the dispersion relation solutions indicate
faster growth for the no wind case. The most likely reason for this
difference is a non-linear surface and first body mode interaction in
the no wind case that is indicated by the observed beat pattern in the
no wind high frequency simulation. Here the different radial structure
of the fundamental and body mode, see Hardee et al.\ (2001), with
comparable wavelength could be responsible for destructive
interference and the reduced transverse velocity growth seen in the no
wind simulation.

The wavelengths and growth/damping lengths normalized to the jet
radius for precession frequencies $\omega 1$, $\omega 2$ and
$\omega 3$ are given in Table 3 for the strongly magnetized simulation
parameters. Weak damping ($wd$) or weak growth ($wg$) indicate a damping
or growth length longer than the grid length in the numerical simulations.
No entry in the growth/damping length column indicates a purely real solution.
\placetable{table 3}
Wavelengths observed in the strongly magnetized simulations are also in
excellent agreement with the theoretically predicted wavelengths.

When no external wind is present the dispersion relation solutions for
the strongly magnetized case show growth of both the S and B1
modes. The growth rate of the S mode is reduced by about 25\% at
frequencies $\omega 1$ and $\omega 2$ and by over a factor of three at
$\omega 3$ when compared to the weakly magnetized case. Thus, we
expect to see the observed increased spatial growth length in the
strongly magnetized no wind simulation when compared to the weakly
magnetized case. The B1 mode now grows at frequencies $\omega 1$ and
$\omega 2$ in addition to growth at frequency $\omega 3$. Growth of
the B1 mode at $\omega 1$ is comparable to growth of the S mode but is
much less than the S mode at $\omega 2$ so would not be expected to
appear in the simulation.  Note that S and B1 mode wavelengths and
growth rates at $\omega 3$ are comparable so we expect an S mode and
B1 mode interaction at a beat wavelength $\lambda_{beat} =
10.6~R_{j}$.  This agrees with a weak beat pattern in $v_{\phi}$ and
$B_r$ (see Figure 7) at $\lambda^{mn}_{beat}(\omega 3)
\sim 10~R_{j}$ seen in the strongly magnetized no wind $\omega 3$
simulation.

The strongly magnetized external wind flow leads to a reduced growth
rate of the $S_{s}$ mode by over an order of magnitude at all
frequencies. The damping rate of the $S_{f}$ mode at frequencies
$\omega \le \omega 2$ is reduced more than the $S_{s}$ growth rate
by about a factor of two.  Note that for the all the other cases the
low frequency growth rate of the $S_{s}$ and damping rate (not
shown) of the $S_{f}$ modes are comparable. At the lower
frequencies, $\omega < \omega 2$ the B1 mode is damped. At
frequencies $\omega 2 < \omega < \omega 3$ the B1 mode is either
purely real or is weakly growing. At the intermediate precession
frequency the strongly magnetized wind case numerical simulation
showed damping and a beat pattern in $B_r$ (see Figure 6) with
$\lambda^{mw}_{beat}(\omega 2) \sim 20~R_{j}$.  The beat pattern can
be understood as interaction between the weakly growing $S_{s}$ and
more weakly damped $S_{f}$ modes with predicted beat wavelength
$\lambda_{beat} = 23.7~R_{j}$.  At the high precession frequency the
strongly magnetized wind case simulation showed slower damping and a
beat pattern in $B_r$ (see Figure 7) with
$\lambda^{mw}_{beat}(\omega 3) \sim 10~R_{j}$.  The beat pattern can
be understood again as interaction between the weakly growing
$S_{s}$ and now more strongly damped $S_{f}$ modes with predicted
beat wavelength $\lambda_{beat} = 10.0~R_{j}$.

At the intermediate and high frequencies the dispersion relation
solutions indicate weak growth of the $S_{s}$ mode but the simulations
suggest damping. However, the radial magnetic field component in
Figures 6 \& 7 show a beat pattern indicating an interaction between
the weakly growing and weakly damped S modes. Normally, a high damping
rate of the $S_{f}$ mode would eliminate observable interaction. We
suggest that the observed damping at $\omega 2$ and the lesser damping
at $\omega 3$ is partially a result of this interaction.  The lesser
damping at $\omega 3$ occurs as interaction is reduced because of the
considerably larger damping rate of the $S_{f}$ mode at the higher
frequency. It is also possible that some of the observed damping is a
result of numerical dissipation given the relatively low numerical
resolution of the simulations.

We can attempt to quantify the growth or damping of the
perturbations seen in the numerical simulations and compare the
observed rates relative to theoretical predictions.  Our estimates of
the growth or damping e-folding length determined from the simulations
are given in Table 4.  The estimates are obtained by comparing a
perturbation amplitude $A_1$, in $v_r$, determined at $z_1$ with a
perturbation amplitude $A_2$ determined at $z_2$. The e-folding growth
or damping length $\ell$ is found from $A_2/A_1 = \exp[(z_2 -
z_1)/\ell]$.  We always choose $z_1 > 3$ to minimize inlet
effects. The range in $z$ over which a growth or damping length was
estimated is included in parentheses following the e-folding
length. An indication of non-linear effects such as an observed beat
pattern or amplitude saturation is indicated by an asterisk.  In these
cases the values provide no more than qualitative guidelines.  The low
frequency result for the no wind case is suspect as a shorter
wavelength perturbation dominates growth at $z > 22$ and the high
frequency cases all involve non-linear mode coupling, indicated by a
beat wavelength, or appear amplitude saturated.  In general, the
estimates vary qualitatively like the theoretical predictions.

We can perform a more quantitative comparison of the moderate
frequency, $\omega = \omega 2$, results with theoretical
growth/damping rate predictions.  The simulation growth lengths are
$1.7 - 2.5$ times longer than theoretically predicted growth
lengths. The weak damping observed in the magnetized wind case at $z
> 26$ when compared to the very weak growth predicted is consistent
with this picture. Part of the difference between theory and
simulation growth lengths can be the result of development of a
shear layer in the simulations where a sharp boundary between spine
and sheath is assumed theoretically. Additional differences can be a
numerical effect resulting from both the narrow width of the
computational domain and the numerical resolution.  The narrow width
of the computational domain means that some interaction with the
domain boundary is unavoidable.  Our relatively low numerical
resolution of 20 computational zones across the jet diameter has the
effect of increasing the numerical viscosity and resistivity.  We do
expect this to affect spatial growth lengths in the manner observed.
We note that resolution studies in 2D RHD simulations performed by
Perucho et al. (2004a, 2004b) suggest that only extremely high
numerical resolution will recover correct quantitative growth or
damping lengths.  Excellent quantitative agreement between
theoretically predicted and our numerically observed wavelengths
indicates that we are not experiencing serious resolution or
boundary effects.

\section{Conclusion}

We have performed numerical simulations of weakly and strongly
magnetized relativistic jets embedded in a weakly and strongly
magnetized stationary or mildly relativistic ($0.5c$) sheath using the
RAISHIN code (Mizuno et al. 2006a). In the numerical simulations a jet
with Lorentz factor $\gamma=2.5$ was precessed to break the initial
equilibrium configuration. Results of the numerical simulations were
compared to theoretical predictions from a normal mode analysis of the
linearized RMHD equations describing a uniform axially magnetized
cylindrical relativistic jet embedded in a uniform axially magnetized
moving sheath.

In the fluid limit the present simulation results confirm earlier
results obtained by Hardee \& Hughes (2003), who found that the
development of sheath flow around a relativistic jet spine explained
the partial stabilization of the jets in their numerical simulations.
Here we confirm this earlier result and have extended the
investigation to the influence of magnetic fields with simulations
specifically designed to test for stabilization of the relativistic
jet spine by strong magnetic fields and a weakly relativistic wind.
The prediction of increased stability of the weakly-magnetized system
with mildly relativistic sheath flow and the stabilization of the
strongly-magnetized system with mildly relativistic sheath flow is
verified by the numerical simulation results.

The simulation results show that theoretically predicted wavelengths
and thus wave speeds are relatively accurate. On the other hand,
growth rates and spatial growth lengths derived from the linearized
equations can only be used to provide broad guidelines to the rate
at which perturbations grow or damp.  Nevertheless, the present
results can be extended to other parameter ranges with reduced
growth occurring as
$$
\left( u_{j}-u_{e}\right)^{2} \rightarrow V_{As}^{2}/\gamma _{j}^{2}\gamma
_{e}^{2}~,
$$
and stabilization occurring when (eq.\ 9)
$$
\left( u_{j}-u_{e}\right)^{2} <  V_{As}^{2}/\gamma _{j}^{2}\gamma
_{e}^{2}~.
$$
In the above expressions
$$
V_{As}^{2}\equiv \left( \gamma _{Aj}^{2}W_{j}+\gamma
_{Ae}^{2}W_{e}\right) \frac{B_{j}^{2}+B_{e}^{2}}{4\pi W_{j}W_{e}}~,
$$
represents a ``surface'' Alfv\'en speed (see eqs. 7 \& 8),
$\gamma_{Aj,e}$ and $\gamma_{j,e}$ are Alfv\'en and flow Lorentz
factors, respectively; $u_{j,e}$, $B_{j,e}$ and $W_{j,e}$ are the
flow, axial magnetic field and enthalpy in the (j) jet or (e) external
sheath.

Formally, the present results and expressions apply only to
magnetic fields parallel to an axial spine-sheath flow in which
conditions within the spine and within the sheath are independent of
radius and the sheath extends to infinity. A rapid decline in
perturbation amplitudes in the sheath as a function of radius is
governed by the Hankel function's radial dependence. This suggests
that the present analysis will provide a reasonable approximation to a
finite sheath provided the sheath is more than about three times the
spine radius in thickness.

In the present regime where flow and magnetic fields are parallel,
current driven (CD) modes are stable (Isotomin \& Pariev 1994,
1996). However, in the strong magnetic field regime we expect the
magnetic fields in realistic jets and sheaths to have a significant
toroidal component and an ordered helical structure. Provided radial
gradients in magnetic fields and other jet spine/sheath properties
are not too large we might expect the present results to remain
approximately valid where $u_{j,e}$ and $B_{j,e}$ refer to the axial
or poloidal velocity and field components only. This conclusion is
suggested by theoretical results, albeit non-relativistic and for a
two dimensional slab jet, indicating that a critical parameter
governing KH stabilization is the difference between the projection
of the velocity shear and the Alfv\'en speed on the normal mode
wavevector (Hardee et al. 1992). In the work presented here magnetic
and flow field are parallel and project equally on the wavevector
which for the helical mode lies at an angle $\theta =
tan^{-1}(1/kR)$ relative to the jet axis.

If flow and magnetic fields are not parallel, the projection of flow
velocity and Alfv\'en velocity on the wavevector is different and this
will modify the stability condition somewhat.  Of course, helical
magnetic fields (Appl \& Camenzind 1992), axially magnetized jet
rotation in the subsonic limit (Bodo et al.\ 1996), and a radially
stratified axial velocity profile (Birkinshaw 1991) do modify the KH
modes. Nevertheless, in the helically twisted magnetic and flow field
regime likely to be relevant to most astrophysical jets where CD modes
are unstable (Lyubarskii 1999), there can be competition between CD
and KH modes. At least in the force-free magnetic field regime, KH
modes can dominate CD modes when both are unstable (Appl 1996).

While the normal Fourier modes, such as the helical mode that we have
considered in this work, are the same in KH and CD regimes, the
conditions for instability, the radial structure, the growth rate and
mode motions are different.  Non-relativistic simulation work (e.g.,
Lery et al.\ 2000; Baty \& Keppens 2003; Nakamura \& Meier 2004)
suggests that CD structure is internal and moves with nearly the jet
speed.  On the other hand, KH structure is surface driven and can move
at speeds much less than the jet speed. These differences may serve to
identify the source of helical or other moving structure on
relativistic jet flows and allow determination of jet properties near
to the central engine required to produce such structure.

\acknowledgments

Y. M. is supported by an appointment of the NASA Postdoctoral Program
at NASA Marshall Space Flight Center, administered by Oak Ridge Associated
Universities through a contract with NASA.  P.\ H. acknowledges partial
support by National Space Science and Technology Center (NSSTC/NASA)
cooperative agreement NCC8-256 and National Science Foundation (NSF) award
AST-0506666 to the University of Alabama. K.\ N. acknowledges partial support
by NSF awards ATM-0100997, INT-9981508, and AST-0506719, and the NASA award
NNG05GK73G to the University of Alabama in Huntsville. The simulations
have been performed on the IBM p690 at the National Center for
Supercomputing Applications (NCSA) which is supported by the NSF and
Altix3700 BX2 at YITP in Kyoto University.

\newpage

\begin{deluxetable}{lcccccc}
\tablecolumns{7}
\tablewidth{0pc}
\tablecaption{Models and Parameters \label{table1}}
\tablehead{
\colhead{Case} & \colhead{$\omega R_{j}/u_{j}$} & \colhead{$u_{e}$} &
 \colhead{$a_{e}/c$} & \colhead{$a_{j}/c$} & \colhead{$v_{Ae}/c$} &
  \colhead{$v_{Aj}/c$}}
\startdata
RHDAn &  0.40  &  0.0  &  0.574  &  0.511  &  0.0682  & 0.064 \\
RHDBn &  0.93  &  0.0  &  0.574  &  0.511  &  0.0682  & 0.064 \\
RHDCn &  2.69  &  0.0  &  0.574  &  0.511  &  0.0682  & 0.064 \\
RHDAw &  0.40  &  0.5  &  0.574  &  0.511  &  0.0682  & 0.064 \\
RHDBw &  0.93  &  0.5  &  0.574  &  0.511  &  0.0682  & 0.064 \\
RHDCw &  2.69  &  0.5  &  0.574  &  0.511  &  0.0682  & 0.064 \\
RMHDBn &  0.93  &  0.0  &  0.30  &  0.226  &  0.56  & 0.45 \\
RMHDCn &  2.69  &  0.0  &  0.30  &  0.226  &  0.56  & 0.45 \\
RMHDBw &  0.93  &  0.5  &  0.30  &  0.226  &  0.56  & 0.45 \\
RMHDCw &  2.69  &  0.5  &  0.30  &  0.226  &  0.56  & 0.45 \\
\enddata
\end{deluxetable}


\begin{deluxetable}{ccccccccc}
\tablewidth{0pt}
\tabletypesize{\small}
\tablecolumns{8}
\tableheadfrac{0.1}
\tablecaption{Wave \& Growth/Damping lengths: Weakly Magnetized
  \label{table 2}}
\tablehead{\colhead{$\omega R_j/u_j$} & \colhead{$\lambda_S(0)$}
  &\colhead{$\ell_S(0)$} &\colhead{$\lambda_B(0)$}
  &\colhead{$\ell_B(0)$} &\colhead{$\lambda_S(u_e)$}
  &\colhead{$\ell_S(u_e)$}
  &\colhead{$\lambda_B(u_e)$} &\colhead{$\ell_B(u_e)$}}
\startdata
0.40 & 13.0 & $6.8g$ & 4.3  & --- & 14.5 & $14.7g$ & 4.4  & $wd$ \\
0.93 & 5.4  & $3.3g$ & 3.2  & --- & 6.2  & $6.9g$  & 3.1  & $wd$ \\
2.69 & $1.83^{\dag}$ & $1.9g$ & $1.65^{\dag}$ & $3.3g$ & 2.14 & $2.7g$  & 1.58
  & $wd$ \\
\enddata
\tablenotetext{\dag}{$(\lambda^{n}_{beat})^{-1} = \lambda_B^{-1} - \lambda_S^{-1}$}
\end{deluxetable}

\begin{deluxetable}{ccccccccc}
\tablewidth{0pt}
\tabletypesize{\small}
\tablecolumns{8}
\tablecaption{Wave \& Growth/Damping lengths: Strongly Magnetized
  \label{table 3}}
\tablehead{\colhead{$\omega R_j/u_j$} & \colhead{$\lambda_S(0)$}
  &\colhead{$\ell_S(0)$} &\colhead{$\lambda_B(0)$}
  &\colhead{$\ell_B(0)$} &\colhead{$\lambda_S(u_e)$}
  &\colhead{$\ell_S(u_e)$}
  &\colhead{$\lambda_B(u_e)$} &\colhead{$\ell_B(u_e)$}}
\startdata
\\0.40 & 12.8 & $9.7g$ & $\left[\begin{array}{c} 4.65 \\ 5.04\end{array}\right]$
& $\left[\begin{array}{c} -- \\ 10.9g \end{array}\right]$
& $\left[\begin{array}{c} 11.7 \\ 15.1 \end{array}\right]$
& $\left[\begin{array}{c} wg \\ wd\end{array}\right]$
& $\left[\begin{array}{c} 4.2 \\ 4.6\end{array}\right]$
& $\left[\begin{array}{c} 38d \\ wd\end{array}\right]$ \\
\\0.93 & 5.4  & $5.7g$ & $\left[\begin{array}{c} 2.38 \\ 3.15\end{array}\right]$
& $\left[\begin{array}{c} 42g \\ -- \end{array}\right]$
& $\left[\begin{array}{c} 5.1 \\ 6.5\end{array}\right]\ddag$
& $\left[\begin{array}{c} 62g \\ wd \end{array}\right]$  & 3.1  & --- \\
\\2.69 & $1.94\dag$ & $6.2g$ & $1.64\dag$ & $7.4g$
& $\left[\begin{array}{c} 1.77 \\ 2.15\end{array}\right]\ddag$
& $\left[\begin{array}{c} wg \\ 19.6d \end{array}\right]$  & 1.69 & $wg$ \\
\enddata
\tablenotetext{\dag}{$(\lambda^{mw}_{beat})^{-1} = \lambda_B^{-1} -\lambda_S^{-1}$}
\tablenotetext{\ddag}{$(\lambda^{mw}_{beat})^{-1} = \lambda_{S_s}^{-1} - \lambda_{S_f}^{-1}$}
\end{deluxetable}

\begin{deluxetable}{ccccc}
\tablecolumns{5}
\tablewidth{0pc}
\tablecaption{Simulation Growth/Damping e-folding lengths \label{table4}}
\tablehead{
\colhead{$\omega R_{j}/u_{j}$} & \colhead{RHDn} &
 \colhead{RHDw} & \colhead{RMHDn} & \colhead{RMHDw} }
\startdata
0.40  &  26 (4-22)  &  30 (3-40)  & ---  & ---  \\
0.93  &  9 (5-22)  &  12 (4-38)  &  14 (5-37)  & $>$ 25$^{\ast}$ wd (26-43) \\
2.69  &  14$^{\ast}$ (11-27)  &  23$^{\ast}$ (7-27)  & $>$ 25$^{\ast}$ wg
(11-25)  & $>$ 19$^{\ast}$ wd (18-35)\\
\enddata
\tablenotetext{\ast}{mode interaction or saturation}
\end{deluxetable}

\end{document}